\documentclass[pss]{wiley2sp}
\usepackage{amsmath}
\usepackage{bm}
\usepackage{w-greek}
\begin{document}

\title{Development of the critical exponent at the antiferromagnetic phase transition of YbRh$_2$Si$_2$ under chemical pressure}

\author{Cornelius Krellner\textsuperscript{\textsf{\bfseries \Ast}},
Christoph Klingner,
Christoph Geibel,
and Frank Steglich}

\mail{e-mail
    \textsf{krellner@cpfs.mpg.de}, Phone: +49 351 4646 2249, Fax: +49 351 4646 2262,
}
\institute{Max Planck Institute for Chemical Physics of Solids, D-01187 Dresden, Germany}


\abstract{%
We present specific-heat measurements in the vicinity of the antiferromagnetic phase transition on single crystals of the alloy Yb(Rh$_{1-x}$Co$_x$)$_2$Si$_2$ for $x\leq 0.38$. This study was motivated by the violation of critical universality in the undoped YbRh$_2$Si$_2$ (Krellner \textit{et al.}, Phys. Rev. Lett. \textbf{102}, 196402) where we have found a large critical exponent $\alpha=0.38$. For Co-doped samples we observe a drastic change of the critical fluctuations resulting in a negative $\alpha$, explainable within the universality classes of phase transitions. The development of $\alpha$ under chemical pressure gives strong indication that the violation of critical universality in YbRh$_2$Si$_2$ is due to the nearby quantum critical point.
}

\maketitle

\section{Introduction}
In recent years, there have been considerable efforts to explore the physics of quantum phase transitions. These transitions are driven by quantum fluctuations in contrast to classical phase transitions which are actuated by temperature. In this context the heavy fermion system YbRh$_2$Si$_2$ was intensively studied, because it is a clean and stoichiometric metal situated on the magnetic side of, but very close to, a quantum critical point (QCP) which can be crossed by applying a tiny magnetic field. Therefore, this system presents both an antiferromagnetic (AFM) phase transition driven by thermal fluctuations as well as pronounced quantum fluctuations (for an experimental and theoretical review see e.g., Ref.~\cite{Gegenwart:2008} and \cite{Misawa:2009}). The recent discoveries of an additional energy scale vanishing at the QCP which does neither correspond to the N\'eel temperature nor to the upper boundary of the Fermi-liquid region  \cite{Gegenwart:2007} and a large critical exponent $\alpha=0.38$ at the AFM phase transition observed in low-temperature specific-heat measurements on a single crystal of superior quality \cite{Krellner:2009} have once again boosted the interest in YbRh$_2$Si$_2$. The latter observation triggered strong theoretical effort to explain the violation of critical universality due to a (quantum) tricritical point \cite{Misawa:2009,Shaginyan:2009}. In this scenario, Misawa \textit{et al.}~\cite{Misawa:2009} proposed the existence of a tricritical point for YbRh$_2$Si$_2$ under pressure and magnetic field at finite temperatures. Experimentally, this part of the phase diagram is easiest to explore using chemical pressure as will be discussed below.\\
The magnetic ordering of YbRh$_2$Si$_2$ ($T_N = 72$\,mK) is stabilized by applying pressure as expected for Yb-Kondo lattice compounds \cite{Plessel:2003}. The complementary method of substituting smaller isoelectronic Co for Rh results in chemical pressure allowing a detailed investigation of the magnetic phase diagram and the physical behavior of the stabilized AFM ordered state. Therefore, a thorough understanding of the physical properties of Yb(Rh$_{1-x}$Co$_x$)$_2$Si$_2$ is of great interest in order to get further insight into the phenomena at the QCP in YbRh$_2$Si$_2$. Very recently, it was shown that for $x=0.07$ the signature of the Kondo breakdown is located within the magnetically ordered phase leading to a detaching of the AFM QCP from the Fermi-surface reconstruction \cite{Friedemann:2009a}. In this contribution, we address the question whether the critical fluctuations around $T_N$ remain anomalous when moving away from the Kondo-breakdown QCP, using positive chemical pressure. To this end, we present detailed specific-heat measurements around $T_N$ for $x=0.12$, 0.27, and 0.38.

\begin{figure}[t]
\includegraphics*[width=\columnwidth]{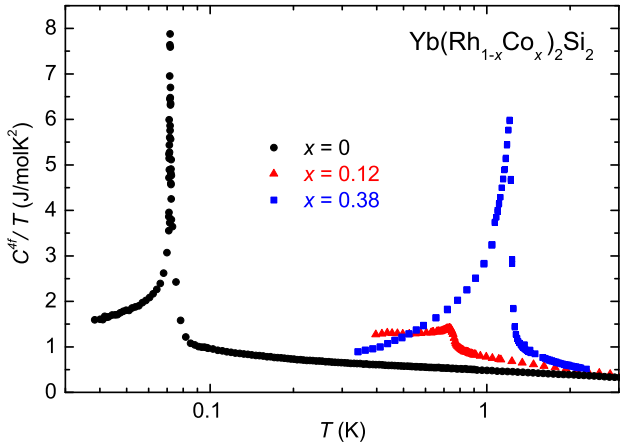}
\caption{\label{fig1}
$4f$ increment to the specific heat plotted as  $C^{4f}/T(T)$ on a logarithmic $T$ scale for three single crystals of the series Yb(Rh$_{1-x}$Co$_x$)$_2$Si$_2$ with $x=0$ (circles, \cite{Krellner:2009}), 0.12 (triangles), and 0.38 (squares), respectively.}
\end{figure}

\section{Experimental}
Single crystals of the alloy series Yb(Rh$_{1-x}$Co$_x$)$_2$Si$_2$ were grown from In flux, analogous to the stoichiometric samples of superior quality \cite{Krellner:2009}. The random substitution of Rh with Co leads to larger disorder compared to YbRh$_2$Si$_2$; however, a comparison with pressure experiments on undoped YbRh$_2$Si$_2$ prove that Co-doping acts mainly as chemical pressure \cite{Friedemann:2009a}. A thorough investigation of the complete doping series, including x-ray diffraction, magnetic susceptibility, electrical resistivity, and specific-heat measurements was performed and will be published separately \cite{Klingner:2009}. The Co-content was accurately determined using energy dispersive x-ray spectra of the polished single crystals.  This real Co-content will be used for $x$ throughout the manuscript. The specific heat for the doped samples was determined with a commercial (Quantum Design) physical property measurement system (PPMS) equipped with an $^3$He-insert, using a standard heat-pulse relaxation technique. The 4$f$ contribution to the specific heat, $C^{4f}$, was obtained by subtracting the non-magnetic one, $C_{\rm Lu}$, from the measured specific heat, $C_{\rm meas}$. $C_{\rm Lu}$ was determined by measuring the specific heat of the non-magnetic reference sample LuRh$_2$Si$_2$ below 10\,K \cite{Ferstl:2007}. Since $C_{\rm Lu}$ at 1\,K contributes only to 1\,\% of $C_{\rm meas}$, the synthesis and measurements of the appropriate reference systems Lu(Rh$_{1-x}$Co$_x$)$_2$Si$_2$ was not necessary.

\section{Results}
In Fig.~\ref{fig1}, the $4f$ increment to the specific heat is plotted as $C^{4f}/T$ on a logarithmic temperature scale for three selected Co-concentrations, $x=0$, 0.12, and 0.38. The anomalies due to the onset of the magnetic order are clearly visible and the type of the anomalies gives already a clue to the development of the critical fluctuations with increasing $x$. Going from $x=0$ to $x=0.12$, we observe a drastic change of the peak form at $T_N$. For the former we observe a very sharp $\lambda$-type peak which leads to a large critical exponent \cite{Krellner:2009}, whereas the latter presents the case of a rather broad mean-field-like anomaly, where it is not possible to extract the critical exponent. For $x=0.38$ we again observe a sharp $\lambda$-type anomaly which we can analyze in terms of critical fluctuations. This development towards a sharper anomaly with increasing $x$ gives evidence that the broad anomaly for $x=0.12$ is not caused by disorder due to the random distribution of the Co-atoms on the Rh side, as the residual resistivity increases together with the doping level from $x=0.12$ to $x=0.38$. It is important to note, that for $0.07\leq x\leq 0.18$ we observe two subsequent magnetic phase transitions \cite{Westerkamp:2008}, whereas for $0.27\leq x\leq 0.38$ we see only one magnetic transition \cite{Klingner:2009} indicating that the magnetic ordering vector might change with increasing $x$.\\
\begin{figure}[t]
\includegraphics*[width=\columnwidth]{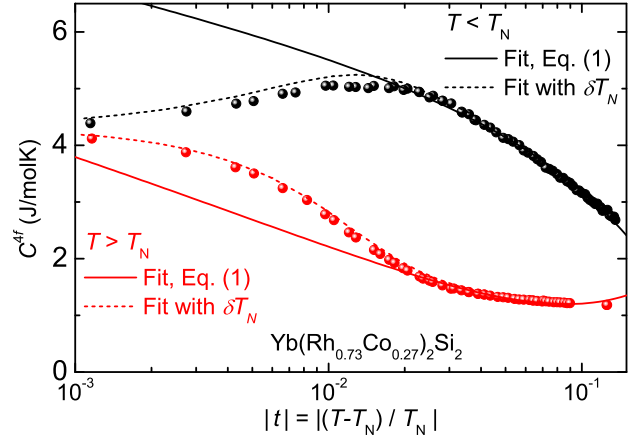}
\caption{\label{fig2}
Specific heat vs reduced temperature close to $T_N$ of Yb(Rh$_{0.73}$Co$_{0.27}$)$_2$Si$_2$. The data below and above $T_N$ can be best fitted with $\alpha^{0.27}=-0.06\pm0.10$ (solid lines). Dotted lines represent the fit including a Gaussian distribution of $T_N$ with $\delta T_N/T_N=6.5\cdot 10^{-3}$ with otherwise identical fit parameters.}
\end{figure}
\begin{figure}[t]
\includegraphics*[width=\columnwidth]{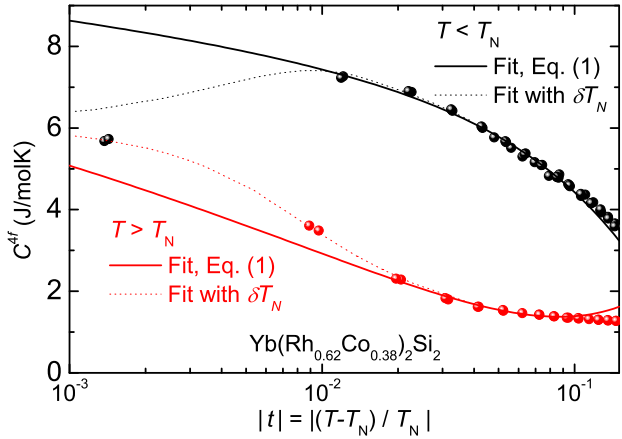}
\caption{\label{fig3}
Specific heat vs reduced temperature close to $T_N$ of Yb(Rh$_{0.62}$Co$_{0.38}$)$_2$Si$_2$. The data below and above $T_N$ can be best fitted with $\alpha=-0.12\pm0.10$ (solid lines). Dotted lines represent the fit including a Gaussian distribution of $T_N$ with $\delta T_N/T_N=5\cdot 10^{-3}$ with otherwise identical fit parameters.}
\end{figure}
\noindent Next, we focus on the analysis of the critical exponent for the two Co-concentrations, $x=0.27$ and 0.38. For $x=0.27$ the peak at $T_N$ looks similar to what is presented for the $x=0.38$ sample \cite{Klingner:2009}. To extract the critical exponent we follow the procedure described in detail in Ref.~\cite{Krellner:2009}. The usual fit function is applied to describe the critical behavior
\begin{equation}\label{eqFit}
C^{\pm}(t) = \frac{A^{\pm}}{\alpha}|t|^{-\alpha} + b + Et\, ,
\end{equation}
with the reduced temperature $t=(T-T_N)/T_N$; $+$($-$) refers to $t>0$ ($t<0$), respectively  \cite{Wosnitza:2007}. The background contribution is approximated by a linear $t$ dependence ($b+Et$) close to $T_N$ \cite{Krellner:2009}. The best fit  for $x=0.27$ reveals a negative critical exponent $\alpha^{0.27}=-0.06\pm0.10$ (Tab.~\ref{TabFit}) and is shown  together with the experimental data in Fig.~\ref{fig2}. To satisfactorily fit the data points for $|t|\leq 0.01$ we have to use a Gaussian distribution of $T_N$ with $\delta T_N=8.4$\,mK to account for the rounding effects, most probably due to the effect of the high doping level. This smearing of $T_N$ leads to a relatively small temperature range, $0.01\leq |t|\leq 0.1$, which determines the critical exponent, severely complicating the analysis and impeding a more accurate determination of $\alpha$. The same analysis was carried out for the measurement of the $x=0.38$ sample and is shown in a similar way in Fig.~\ref{fig3}. The curves exhibit a comparable overall $t$-dependence like the $x=0.27$ data, giving again a negative critical exponent $\alpha^{0.38}=-0.12\pm0.10$. Here, we have to use a slightly smaller distribution of $T_N$ with $\delta T_N/T_N=5\cdot 10^{-3}$ consistent with a lower residual resistivity in the $x=0.38$ crystal  \cite{Klingner:2009}.\\
\begin{table}[tb]
\caption{\label{TabFit} Parameters obtained from the fits of the specific-heat around $T_N$ for $x=0$ \cite{Krellner:2009}, $x=0.27$, and $x=0.38$.}
\begin{tabular}{ccccc} \hline
$x$  	&	$T_N$ 	& $\delta T_N/T_N$	 & $A^{+}/A^{-}$& $\alpha$ 	\\
		&	(K)		&		$(10^{-3})$	&		&  			\\	\hline
0		& 	$0.072$	& 	0.3				&	 0.6(1)			&	$+0.38(3)$	\\
0.27	& 	$1.298$	& 	6.5				&	 1.4(5)			&	$-0.06(10)$	\\
0.38	& 	$1.223$	& 	5				&	 2.2(5)			&	$-0.12(10)$	\\ \hline \hline
\end{tabular}
\end{table}
\noindent In Tab.~\ref{TabFit} we summarize the parameters obtained from the analysis of the critical exponent for the doped samples presented here and compare them with the ones obtained for the undoped compound \cite{Krellner:2009}. It is obvious that there is a drastic change of the critical exponent going from $\alpha=0.38(3)$ for $x=0$ to $\alpha=-0.12(10)$ for $x=0.38$. The latter value can be explained in terms of the classical universality classes in the theory of phase transitions, for which $-0.133(5)\leq \alpha \leq +0.110(1)$ generally holds true.  However, the low experimental accuracy of the determined exponent is not sufficient to distinguish between the two applicable symmetry classes \cite{Wosnitza:2007}, namely the 3D-Heisenberg model $[\alpha_{3D,H}=-0.133(5)]$ or the 3D-XY model $[\alpha_{3D,XY}=-0.015(1)]$. More important is the clear development of $\alpha$ from an unconventional value to a conventional one when moving away from the Kondo-breakdown QCP by chemical pressure which gives a first indication that the violation of critical universality in YbRh$_2$Si$_2$ is due to this nearby unconventional QCP which may substantially influence the spatial fluctuations of the classical order parameter.\\
Furthermore, our result of a conventional critical exponent for YbRh$_2$Si$_2$ under chemical pressure is in contradiction to what is expected in the quantum-tricritical-point scenario \cite{Misawa:2009}, where Misawa \textit{et al.} suggest a stabilized tricritical point at finite temperatures in the pressurized YbRh$_2$Si$_2$. In this case one would expect that the critical exponent becomes even larger for YbRh$_2$Si$_2$ under chemical pressure, as the theoretical value at a classical tricritical point is $\alpha=0.5$, just opposite to what we have observed.

\section{Conclusions}
In conclusion, we have presented specific-heat measurements of Co-doped YbRh$_2$Si$_2$ in the vicinity of $T_N$ and studied the development of the critical exponent. We found that the critical fluctuations change drastically with doping. The broad mean-field-type anomaly at $T_N$ for $0.07\leq x \leq 0.18$  prevents the determination of the critical exponent, but for $x=0.27$ and $0.38$ we found a conventional critical exponent of $\alpha=-0.06(10)$ and $-0.12(10)$, respectively. The  development of $\alpha$ from an unconventional value to a conventional one when moving away from the local QCP by chemical pressure gives strong indication  that the violation of critical universality in YbRh$_2$Si$_2$ is due to this nearby QCP.

\begin{acknowledgement}
We acknowledge fruitful discussions with M. Brando, P. Gegenwart, S. Kirchner, M. Nicklas, Q. Si, and J. Wosnitza. 
The Deutsche Forschungsgemeinschaft (SFB 463, Research Unit 960) is acknowledged for financial support.
\end{acknowledgement}

\providecommand{\WileyBibTextsc}{}
\let\textsc\WileyBibTextsc
\providecommand{\othercit}{}
\providecommand{\jr}[1]{#1}
\providecommand{\etal}{~et~al.}

\providecommand{\WileyBibTextsc}{}
\let\textsc\WileyBibTextsc
\providecommand{\othercit}{}
\providecommand{\jr}[1]{#1}
\providecommand{\etal}{~et~al.}

\end{document}